\documentclass[aps,prb,preprint,groupedaddress,floatfix]{revtex4}
\usepackage{graphicx,amssymb}

\begin{document}

\title{Stabilization of nanobubbles under hydrophobic confinement.} 
\author{Caroline Desgranges and Jerome Delhommelle}
\affiliation{Department of Chemistry, University of North Dakota, Grand Forks ND 58202}
\date{\today}

\begin{abstract}
The lifetime of nanobubbles can exceed by more than 10 orders of magnitude the theoretical expectation, which predicts an almost immediate dissolution due to their very high Laplace internal pressure. This makes nanobubbles promising candidates for energy applications, as high-pressure nanoreactors in fuel cells, and for gas delivery in biological systems. Here, we use molecular simulation to shed light on the formation and stabilization of nanobubbles in carbon nanotubes. Using an entropic order parameter, we elucidate the nucleation pathway and determine its free energy. We identify a critical volume for which the existence of nanobubbles is thermodynamically favored, with a flat free energy profile around this critical volume, and mechanically favored, since the fluid pressure along the nanotube axis is positive at this juncture. The stabilization process is assisted by the hydrophobic nature of the nanotube and by the formation of strong hydrogen bonds at the interface.
\end{abstract}

\maketitle

\section{Introduction}

The formation of nanobubbles, {\it i.e.}, of nanoscopic gaseous domains within liquids, has drawn considerable interest in recent years~\cite{Lohse,Maheshwari,Seddon1,Seddon2,Parker,Ishida}. One of the most striking and intriguing features of nanobubbles is their unexpected stability, for instance, of up to two weeks for a $50$ nm nanobubble~\cite{Ohgaki}, despite the theoretical expectation that these domains should be unstable and dissolve as a result of their high internal Laplace pressure. This amazing property of nanobubbles has thus made them emerge as promising candidates for many applications, {\it e.g.}, as high pressure nanoreactors with enhanced reaction kinetics~\cite{Svetovoy} that could be employed within fuel cells without the need for expensive catalysts. Other potential applications include their use as ultrasound contrast agents, as transport for gas delivery to membranes and cells~\cite{Dzubiella} which could have effects on transmembrane proteins and on membrane structures, in turn, modifying cell function and promoting biological functions. Furthermore, nanobubbles can serve as a seed for the growth of larger bubbles, responsible for decompression sickness~\cite{Craig}. Other recent developments in nanotechnology involve the role of nanobubbles to promote the motion of nanomotors by bubble propulsion for nanomedicine applications~\cite{Wang}. Several mechanisms have been proposed to account for the remarkable stability of nanobubbles, either through the assistance of nearby substrates in the case of surface nanobubbles~\cite{Zhang,Brenner,Weijs} or through the adsorption of ions on the outside of bulk nanobubbles~\cite{Ohgaki}. However an understanding of nanobubble formation at the molecular level still remains elusive as none of these mechanisms fully account for the stabilization of such objects.

From an experimental standpoint, surface bubbles can be generated, for instance, through electrolysis~\cite{Zhang2}, while nanobubble solutions are induced mechanically by passing a solution through a small space~\cite{Ohgaki}. On the basis of these experiments, several possible explanations have been proposed to account for the stabilization of nanobubbles. For instance, in the case of surface nanobubbles, it was suggested that hydrophobic surfaces play a key role in the stabilization process~\cite{Zhang,Tan} as they tend to adsorb a vapor phase, which, in turn, provides a mechanism for the stabilization process through a thermodynamic equilibrium~\cite{Brenner,Weijs}. In the case of bulk nanobubbles, the stabilization process seems to be connected to the formation of a strong hydrogen bond network at the interface, as evidenced by the results from attenuated total reflectance infrared spectroscopy~\cite{Ohgaki}. Recent work has also shown that the evaporation of water confined between hydrophobic planes could result in the formation of bubbles~\cite{Remsing,vembanur2013thermodynamics,Altabet}. Here, we use molecular simulation to shed light on the stabilization of nanobubbles within carbon nanotubes filled with water, a system that has drawn considerable interest given its potential applications~\cite{Waghe,Zambrano,Nosonovsky,Rossi,Tas,Gubbins,Alexiadis,Pascal,Bernardina,kannam2017modeling}. The simulation allows us to identify a mechanism starting with the initial formation of small surface nanobubbles that coalesce to yield a large gaseous nanodomain stabilized by the adsorption of a vapor phase close to the hydrophobic walls. Our results also provide a rationale at the molecular level, for the prolonged stability of surface nanobubbles. More specifically, our analysis leads to the determination of a critical volume for the nanobubble, which becomes thermodynamically favored, since the free energy profile is remarkably flat at this point, as well as mechanically favored, since the pressure, in the direction parallel to the nanotube axis, reaches a positive value for a nanobubble of this size. This also suggests a new way of preparing surface nanobubbles through the choice of the thermodynamic conditions, instead of using electrolysis or some mechanical means. This new preparation can also be readily applied to other single component systems as well as to mixtures, which are of particular interest in energy applications and fuel cells.

The paper is organized as follows. In the next section, we present the simulation approach used in this work, known as the $\mu VT-S$~method~\cite{Desgran1}, that allows us to compute the free energy barrier associated with the formation of the nanobubbles and to identify the bubble nucleation pathway. We also provide a brief account of the force fields used in this work, specifically the $SPC/E$ model~\cite{Berendsen} for water and how interactions between water and carbon nanotubes are calculated~\cite{Pascal}.  We then discuss how we identify the thermodynamic conditions leading to bubble formation for water confined in the nanotubes and determine the corresponding free energy profiles. In particular, we analyze, at the molecular level, the mechanism underlying the nucleation of nanobubbles and examine the thermodynamic and mechanical factors favoring the formation and stability of water nanobubbles. We finally draw the main conclusions of this work in the last section.

\section{Simulation method and models.}

\subsection{Determining the free energy profile for nanobubble formation}

To understand the formation of the metastable states spanning the nanobubble formation pathway~\cite{Debenedetti}, we employ a molecular Monte Carlo (MC) simulation method that allows for the sampling of rare events. Here, we achieve this by using the recently developed $\mu VT-S$ method~\cite{Desgran1,desgranges2016free1,desgranges2016free2}, which employs an umbrella sampling (US) potential~\cite{Torrie,tenWolde,PRL,Remsing,Monson,Desgranges3,Desgranges4,desgranges2018unusual,desgranges2018crystal} to overcome the free energy barrier associated with bubble formation and to allow us to sample the entire pathway. This US potential is function of an order parameter that characterizes the structural changes taking place within the nanoconfined fluid, as the formation of the nanobubble proceeds. 
In this work, we choose an entropic order parameter, calculated from the Helmholtz free energy and internal energy of the confined fluid~\cite{Desgran1,Waghe}. Specifically, for each configuration $i$ sampled during the MC simulations, we calculate the value of the order parameter $S_i=(U_i-A)/T$, in which $U_i$ is the sum of the potential energy for configuration $i$ and of the kinetic energy ($3Nk_BT$ for $N$ water molecules) and $A$ is the Helmholtz energy. We add that the kinetic contribution is of $3k_BT$ per molecule, since we use a rigid model for water with six total degrees of freedom, three for the translations and three for the rotations. In line with prior work by Waghe, Hummer and Rasaiah on the entropy of water in CNTs~\cite{Waghe}, $pV$ is found to be small when compared to the values taken by $\mu$, leading us to take $A$ to be equal to $N\mu$ (for a system of $N$ water molecules) in the equation for the entropic order parameter. The umbrella sampling potential $V_{US}=(S_i-S_0)^2/2$ is a harmonic function, function of the difference between the value taken by the entropic order parameter $S_i$ and the target value for the entropy $S_0$. From a practical standpoint, we typically perform 60 $\mu VT-S$ simulations, with decreasing values for the imposed value for the overall entropy of the system, to sample the entire nucleation pathway. This defines $60$ overlapping windows, over which statistics are collected and the free energy profile for the nucleation process can be calculated using the conventional method for US simulations~\cite{Torrie}. 

\subsection{Force Fields}

We use the $SPC/E$ force field~\cite{Berendsen} for water, which models well the thermodynamics of the vapor-liquid equilibrium for water~\cite{Boulougouris,errington1998molecular,vega2011simulating,NIST,desgranges2016evaluation,desgranges2016ideality,Desgranges5}. In particular, prior simulations, performed using the Expanded Wang Landau simulation method~\cite{desgranges2012evaluation}, have allowed us to identify the chemical potential for the $SPC/E$ model at the vapor-liquid coexistence~\cite{Desgranges5}. The thermodynamic conditions used in $\mu VT-S$ simulations, with $\mu$ ranging from -4120~kJ/kg to -4160~kJ/kg for $T=500$~K, are chosen close to the coexistence for bulk water. Prior work has shown that the surface tension $\gamma$ for the $SPC/E$ model~\cite{Vega} at $500$~K is of $25.9$~$mJ/m^2$, in reasonable agreement with the experimental value of $31.61$~$mJ/m^2$. This means that, under such conditions, the Laplace (internal) pressure exceeds the fluid pressure by $2\gamma/R$, which corresponds, for a nanobubble of radius $R=10-50$~\AA, to an excess pressure ranging from $10$~MPa to $50$~MPa range. Such nanobubbles should dissolve quickly in the absence of any surface-mediated stabilization mechanism~\cite{Seddon1}. To elucidate this stabilization mechanism, we simulate the nucleation of nanobubbles in water confined in carbon nanotubes~\cite{Zambrano,Nosonovsky,Rossi,Tas,Gubbins,Alexiadis,Pascal,Bernardina} (CNTs) with an armchair configuration $(n,n)$, and use the water-CNTs interaction parameters of Goddard {\it et al.}~\cite{Pascal}. To determine the impact of the nanoscopic confinement on the nucleation process, we systematically vary the CNT diameter by increasing $n$ from $14$ to $26$. 

\subsection{Technical details}

The $\mu VT-S$ simulations are implemented within a Monte Carlo (MC) framework, with the usual MC moves for the grand-canonical ensemble corresponding either to a translation of a randomly chosen water molecule, to the rotation of a random water molecule, and to the insertion or to the deletion of a molecule. 60 simulations are typically carried out to sample the entire pathway connecting the nanoconfined vapor phase to the nanoconfined liquid. For each window, we carry out an equilibration run of $5\times10^6$ MC steps, before a production run of $10\times10^6$ MC steps is performed over which averages and histograms are collected. A cutoff of $15$\AA~ is used for the Lennard-Jones interactions, while Ewald sums are used for the electrostatic interactions following the method of Yeh and Berkowitz~\cite{yeh1999ewald}. In the rest of this work, we work with the usual system of reduced units~\cite{Allen} for the order parameter $S^*$, with the unit for energy being defined by $\epsilon_{SPC/E}/k_B$, the unit for length by $\sigma_{SPC/E}$ and the unit for mass by the molar mass of water. We also perform several different structural analyses to characterize the nanoconfined fluid properties, as well as the features of the developing nanobubble. First, we determine the void fraction within the nanotube by dividing the volume inside the nanotube into small volume elements, with a 0.5$\sigma_{SPC/E}$~interval along the nanotube axis, and a 0.05$\sigma_{SPC/E}$~interval in the radial direction. This allows us to calculate the local density of confined water from the configurations generated during the simulations, with void volume elements identified as containing zero water molecules. Second, we evaluate the number of hydrogen bonds $N_{HB}$, in which each water molecule is involved, according to the usual geometric criterion~\cite{Kuffel,maerzke2010effects}. We also keep track, for each molecule $i$, of the number of neighboring water molecules, $N_n(i)$, for which the $O-O$ distance between two water molecules must be less than 3.5~\AA. Finally, we adapt the spatial analysis of Levitt and Sharon~\cite{Levitt}, devised to analyze the properties of water molecules in the vicinity of proteins, to characterize the behavior of water molecules in CNTs. For this purpose, we define 4 classes of water molecules, denoted by roman numerals from I to IV. Classes I and II include water molecules close to the CNT surface (less than 3.6~\AA~away from the $C$ atoms of the CNT), with class I molecules having few nearest neighbors ($N_n$ < 2.8) than class II molecules. Classes III and IV correspond to water molecules located deeper inside the CNT (more than 3.6~\AA~away from the $C$ atoms of the CNT), with class III molecules having fewer nearest neighbors ($N_n$ < 3.4) than class IV molecules.

\section{Results and discussions}

\begin{figure}
\includegraphics*[width=10cm]{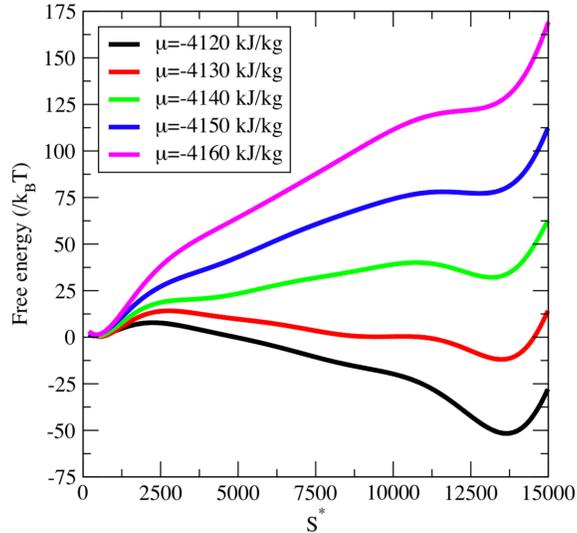}
\caption{Water in $(20,20)$ CNT: Free energy profiles for $\mu$ ranging from -4120~kJ/kg to -4160~kJ/kg as a function of the entropic order parameter}
\label{Fig2020}
\end{figure}

The free energy profiles, obtained for $\mu$ ranging from -4120~kJ/kg to -4160~kJ/kg, are shown in Fig.~\ref{Fig2020} on the example of the $(20,20)$ CNT. In all cases, the free energy profile connects a first minimum, reached for a low value of the reaction coodinate $S^*$ around 200, to a second minimum, for a value of $S^*$ around 13500. The order parameter $S^*$ is extensive~\cite{Desgranges5}, since it is calculated as the total entropy of nanoconfined water, and takes into account the increase in $N$, the number of water molecules confined in the nanotube (note that the total entropy is not strictly proportional to $N$, since it is function of the number of molecules, as well as of the amount of order within the confined fluid). As a result, the first minimum for a low total $S^*$ is associated with configurations with low $N$ values, corresponding to a vapor phase adsorbed in the CNT. On the other hand, the second minimum, reached for a high total $S^*$ and large $N$ values, corresponds to a nanoconfined liquid phase. For instance, for $\mu=-4140$~kJ/kg and a CNT length of about 100~\AA, we have an average $<N>=17.5$~molecules at the first minimum, and $<N>=1077.7$~molecules at the second minimum. Comparing the free energies associated with each minimum also allows us to identify the conditions for which the nanoconfined vapor is more stable, for $\mu=-4140$~kJ/kg and above, and for which the nanoconfined liquid is more stable, for $\mu=-4130$~kJ/kg and below. We finally add that the free energy profiles, obtained here for water in CNTs using the $\mu VT-S$ approach, are consistent with those found for the evaporation and condensation of water confined between hydrophobic planes.~\cite{Remsing,vembanur2013thermodynamics,Altabet}

\begin{figure}
\includegraphics*[width=8cm]{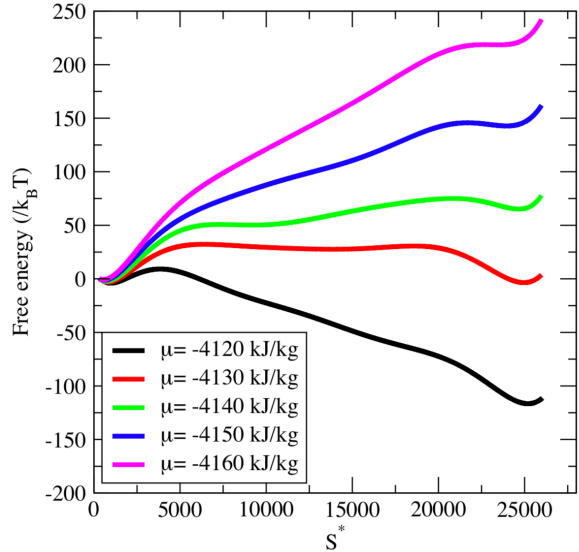}
\caption{Water in $(26,26)$ CNT: Free energy profiles for $\mu$ ranging from -4120~kJ/kg to -4160~kJ/kg as a function of the entropic order parameter}
\label{Fig2626}
\end{figure}

We now examine the impact of the diameter of the CNT on the free energy. The profiles obtained in the case of the $(26,26)$ CNT are shown in Fig.~\ref{Fig2626}. The features are qualitatively similar to those found for the $(20,20)$ CNT, with all profiles exhibiting two minima corresponding to the nanoconfined vapor and liquid. A notable difference is the wider range of $S^*$ spanned during the process. This is due to the fact that the $(26,26)$ CNT has a larger diameter and can accomodate almost twice as many water molecules as the  $(20,20)$ CNT for the same length, resulting in much larger $S^*$ values. Another consequence of the increase in the diameter is the increased sharpness in the phase transition. This can be seen from the difference in free energy between the nanoconfined vapor and liquid phases, which goes from about -120$k_BT$ ($\mu$=-4120~kJ/kg) to 65$k_BT$ ($\mu$=-4140~kJ/kg) for the $(26,26)$ CNT. Over the same range of $\mu$, this difference only increases from about -50$k_BT$  to 30$k_BT$ for the $(20,20)$ CNT.

\begin{figure}
\includegraphics*[width=10cm]{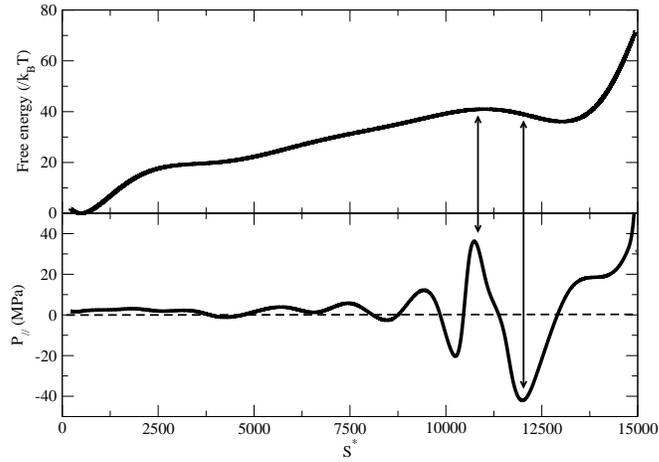}(a)
\includegraphics*[width=10cm]{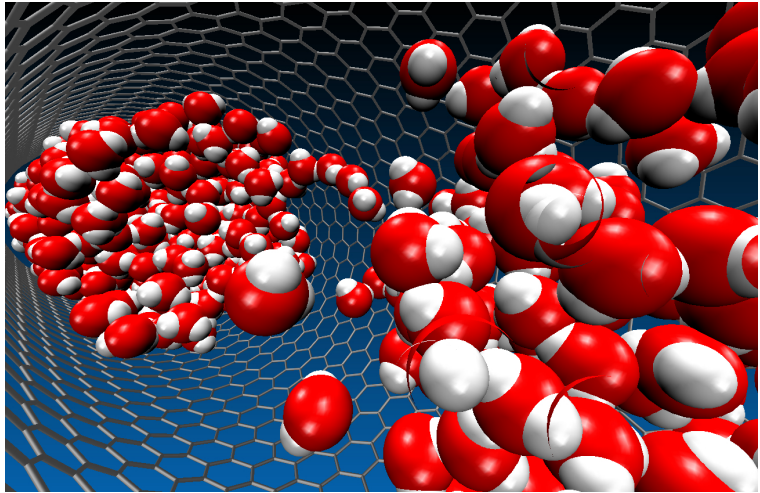}(b)
\caption{(a) Correlation between the free energy profile (top) associated with the dewetting process, from right to left, and the variation of the parallel pressure ($P_{//}$) of water in CNT $(20,20)$ for $\mu=-4140$~kJ/kg. The arrow on the right indicates the onset of the nanobubble nucleation process, with a sharp increase in free energy, and a sharp drop in $P_{//}$ which becomes negative. The second arrow corresponds to a flat part of the free energy profile, associated with a positive value for $P_{//}$ and, at the molecular level, with the nanobubble shown in (b). }\label{Fig1}
\end{figure}

\begin{figure}
\includegraphics*[width=14cm]{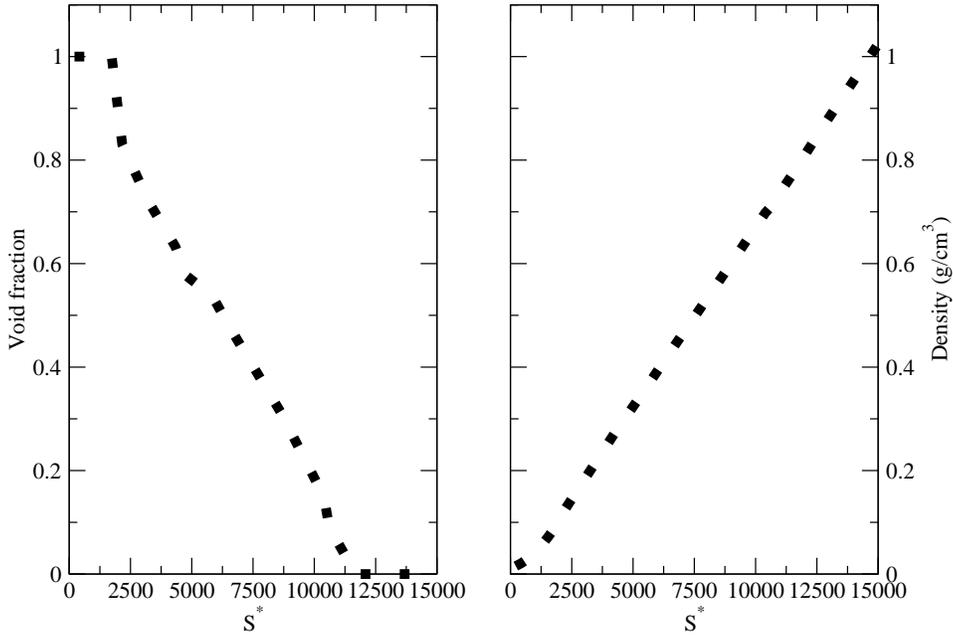}
\caption{(Left) Void fraction in CNT $(20,20)$ during the nanobubble nucleation process at $\mu=-4140$~kJ/kg, and (Right) Density change in the nanoconfined fluid along the nucleation pathway. }\label{Fig2}
\end{figure}

We now focus on the case of the $(20,20)$ CNT and examine more closely the properties of nanoconfined water along the entropic pathway for $\mu=-4140$~kJ/kg. The free energy profile for the nanobubble nucleation process is shown in Fig.~\ref{Fig1}. Since the formation of a bubble in a capillary generally corresponds to the onset of a negative pressure in the system, as shown e.g. in acoustic experiments~\cite{Caupin} and simulations~\cite{Giovambattista}, we also determine the fluid pressure along the nanotube axis, $P_{//}$, through the virial expression, and plot its variation along the nucleation pathway. The nucleation pathway starts from the right hand side of the plot, i.e. from a completely filled CNT, with a high water loading and thus a high total entropy $S^*$, and a positive value for $P_{//}$. At this point, the system is a metastable nanoconfined liquid, associated with a local minimum in free energy reached for $S^*$ around $13500$. We carry out a structural analysis to confirm the nature of the confined fluid. Fig.~\ref{Fig2} shows that the void fraction is equal to $0$ for this value of $S^*$ and that the density of water in the CNT is of $0.94$~$g/cm^3$, which is typical of nanoconfined water. As $S^*$ decreases, the nanobubble nucleation process starts to take place with the formation of very small cavities close to the hydrophobic surface of the CNT. This is the first part of the nucleation process, which is characterized by an increase in the free energy of the nanoconfined fluid, and by a pronounced dip in $P_{//}$, which becomes negative (see, {\it e.g.}, Fig.~\ref{Fig1}a for $S^*=12000$). During this stage, the cavities that form close to the surface are very small and have a very large internal pressure. This in turn results in a system that is both mechanically unstable, with a strongly negative $P_{//}$, and thermodynamically unstable as the free energy profile exhibits a significant slope during this stage. The formation of this cavities can also be monitored through the steady, albeit slow, increase in the void fraction, and through the decrease of the water loading seen on the right panel of Fig.~\ref{Fig2}. 

\begin{figure}
\includegraphics*[width=12cm]{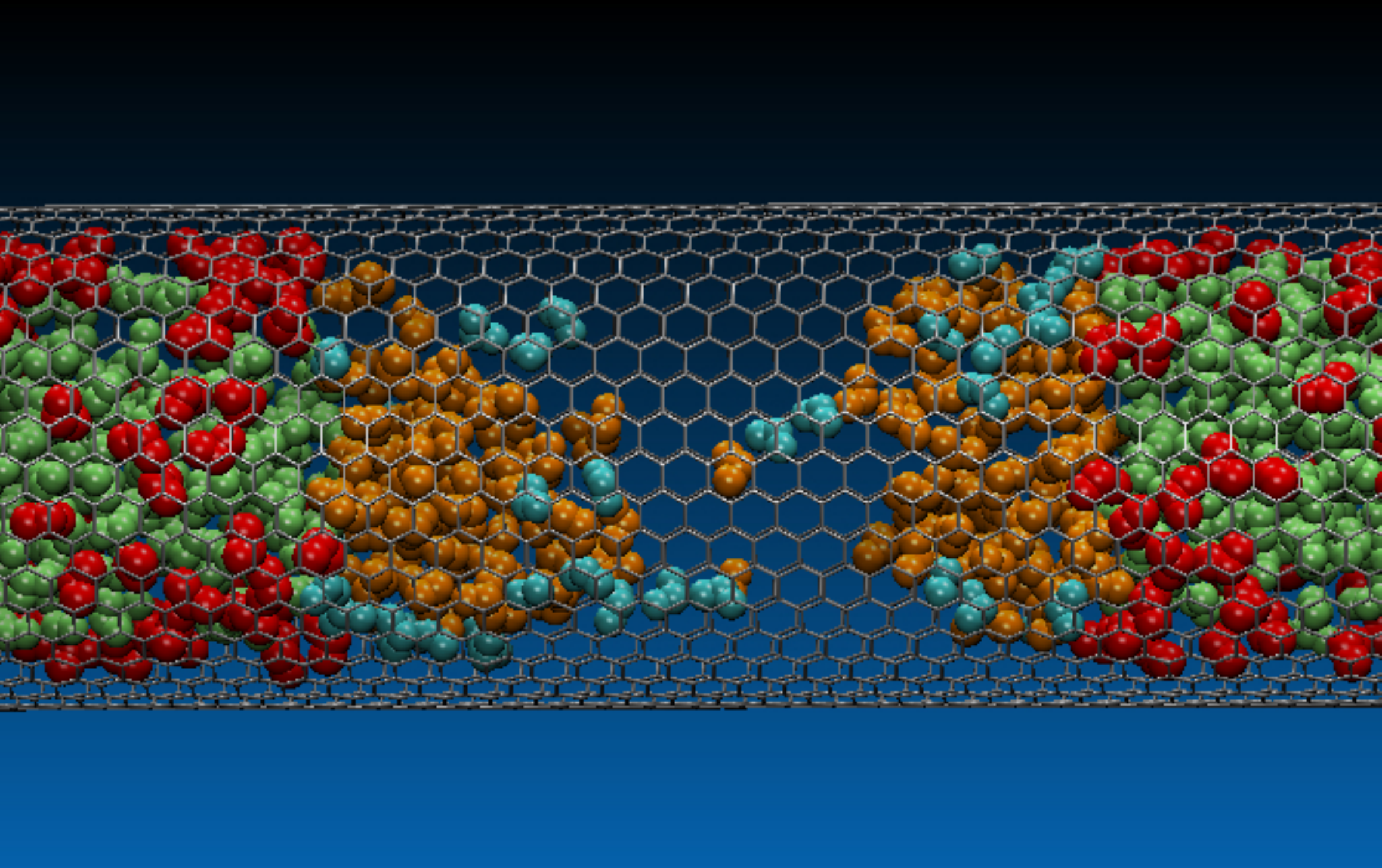}(a)
\includegraphics*[width=12cm]{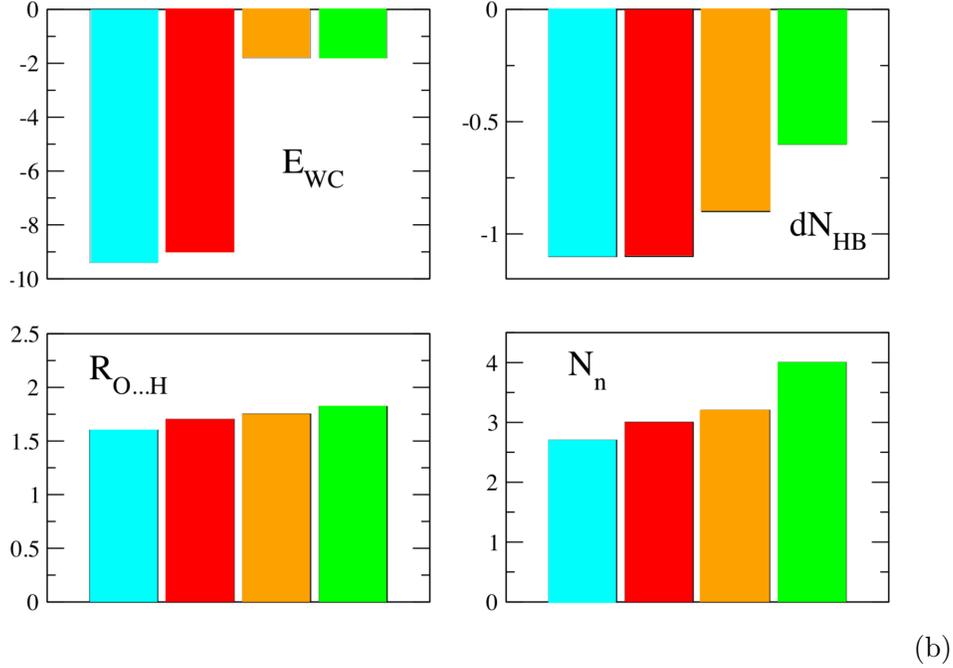}(b)
\caption{Definition for the four classes of water during the nanobubble nucleation process (CNT $(20,20)$ for $S^*=10600$). (a) Spatial distributation of Class I (in cyan), Class II (in red), Class III (in orange) and Class IV (in green). (b) Energetic and structural features of the four classes with (top left corner) the water-CNT interaction energy ($E_{WC}$ is given in units reduced with respect to $\epsilon~SPC/E$), (top right corner) the number of hydrogen bonds per water molecule relative to the bulk ($dN_{HB}$), (bottom left corner), the $O \cdots H$ distance in a hydrogen bond $O-H \cdots O$ ($R_{O \cdots H}$) and (bottom right corner) the number of neighboring water molecules in a shell of 3.5~\AA~radius ($N_n$).}\label{Fig3}
\end{figure}

\begin{figure}
\includegraphics*[width=10cm]{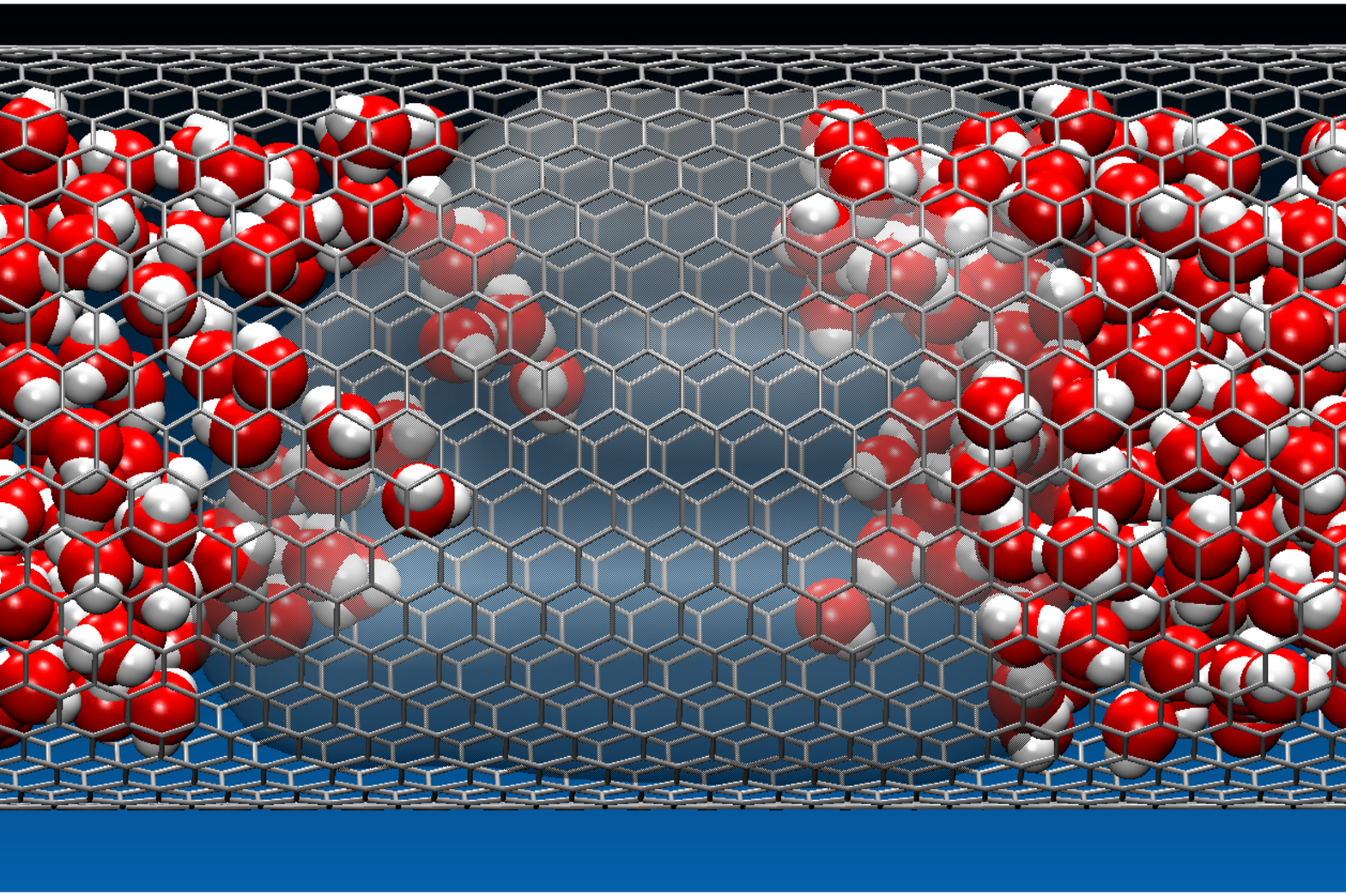}(a)
\includegraphics*[width=10cm]{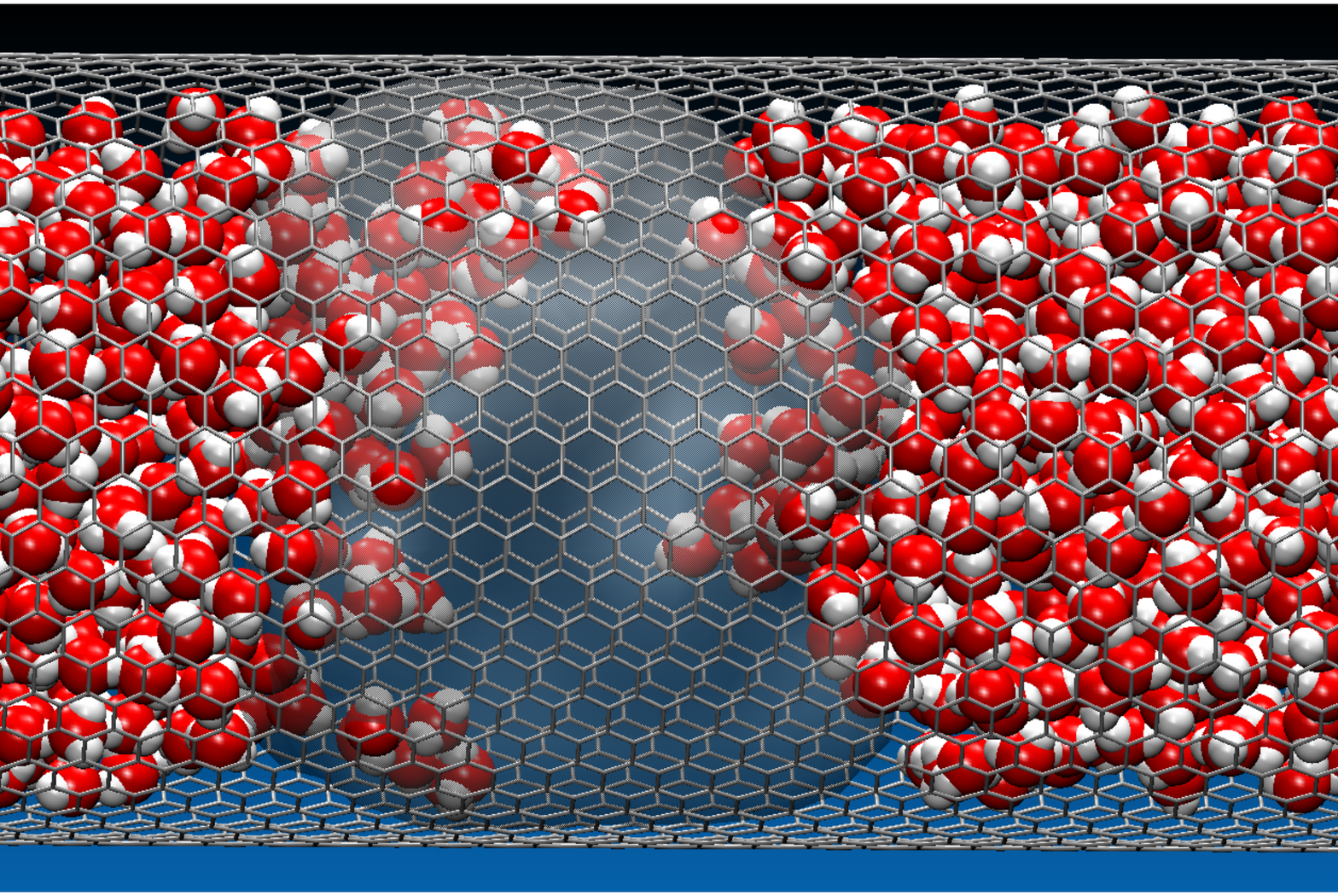}(b)
\caption{Nanobubble formation in $(20,20)$ CNT (a) and in $(26,26)$ CNT (b) for $\mu=-4140$~kJ/kg. Nanobubbles, shown here as a translucent surface, are taken when both favored thermodynamically and mechanically, as indicated in Fig. 1a. Nanobubbles critical volumes are of $2225$~\AA$^3$ for CNT $(20,20)$ and of $5371$~\AA$^3$ for CNT $(26,26)$.}\label{Fig4}
\end{figure}

As $S^*$ further decreases, the void fraction undergoes a more rapid increase, as the cavities start to coalesce. Then, as coalescence further advances, a thorough reorganization takes place within the fluid with the formation of a nanobubble across the nanotube. The snapshot, plotted in Fig.~\ref{Fig1}b, is obtained when $S^*$ reaches $10600$. It shows a typical configuration of the nanoconfined liquid, showing that a nanobubble, surrounded by the nanoconfined liquid has nucleated. At this point, the free energy profile reaches a maximum, indicating that the free energy of nucleation of the nanobubble is of $10\pm1$~$k_BT$, and that the critical volume for the nanobubble is of $2225\pm150$~\AA$^3$. Furthermore, the plot shows that the free energy profile becomes flat, while $P_{//}$ becomes positive again. The combination of these two factors results in a stabilization of the nanobubble, both mechanically since the fluid pressure is positive again, and thermodynamically, since the system is on a flat part of the free energy profile. This implies that the nanobubble so obtained can remain metastable over a prolonged period of time, thereby providing insight into the unexpected stabilization of a nanobubble. Then, as $S^*$ further decreases, the free energy profile leaves the plateau and starts to decrease again. This decrease in free energy occurs concomitantly with a decrease in $P_{//}$, as well as a continued increase in void fraction and a decrease in the density of nanoconfined water. This indicates that nanobubbles, with a volume exceeding the critical volume, start to spontaneously grow and take over the system. The free energy profile then continues to decrease, while $P_{//}$ converges towards the pressure of a vapor phase of water adsorbed in the nanotube. At this stage, the free energy profile reaches its minimum, indicating that the system has reached its stable phase, the nanoconfined vapor, which marks the end of the pathway for $S^*$ around $200$. We finally add that the free energy profile obtained in this work is consistent with those found for the dewetting process under nanoconfinement using umbrella sampling simulations~\cite{Remsing,vembanur2013thermodynamics} as well as forward-flux sampling simulations~\cite{Altabet} for rigid confining plates.

This now prompts the question of identifying the structure and organization, at the molecular level, accounting for this phenomenon. Experiments, as well as theoretical approaches, have suggested that a specific mechanism and microscopic organization takes place at the vapor-liquid interface to account for the unexpected stability of the nanobubbles~\cite{Lohse,Seddon1,Weijs}. We therefore carry out an analysis of the interaction energy between water molecules and the CNT ($E_{W-CNT}$), the number of hydrogen bonds per molecule relative to the number of hydrogen bond for the bulk~\cite{Kuffel,maerzke2010effects} ($dN_{HB}$), the $O\cdots H$ distance in a hydrogen bond $O-H \cdots O$ and the number of neighboring water molecules within a $3.5$~\AA~shell ($N_n$). This analysis allows us to identify the four different classes of water molecules, shown in Fig.~\ref{Fig3}(a) on a snapshot, together with the corresponding data in Fig.~\ref{Fig3}(b). The nanobubble is composed of Class I and Class III molecules, with Class I molecules being located close to the CNT surface and Class III molecules located at the surface of the nanobubble, away from the CNT. The rest of nanoconfined water is composed of Class II molecules (water molecules not involved in the nanobubble and close to the CNT) and of Class IV molecules (away from both the nanobubble surface and the CNT). Focusing first on the molecules close to the CNT, Class I molecules differ from Class II molecules by the smaller values of $N_n$ observed for Class I. Low values of $N_n$ corresponds to a reduced number of neighboring water molecules for Class I molecules, and thus, lead us to assign a vapor-like character to Class I molecules. Our results point to having a part of the nanobubble adsorbed at the three-phase contact line and stabilized by the attractive Lennard-Jones interaction between the Class I water molecules and the $C$ atoms of the CNT. This is in line with the theoretical expectations and experimental observations for nanobubble stability~\cite{Lohse}. In particular, hydrophobic surfaces, that repels liquid water and adsorbs vapor-like molecules, have been found to reduce the surface area through which outfluxing takes place and, in turn, to stabilize nanobubbles~\cite{Seddon1}. While the simulation results generally confirm the expected role played by the hydrophobic surface of the CNT~\cite{Lohse,Seddon1}, further work is needed to study systematically how the stability of the nanobubble can be further controlled by fine tuning the water-surface interactions. We now turn to the comparison between Class III and Class IV molecules, that are both located away from the CNT. Looking at the number of hydrogen bonds per molecule, we find that Class III molecules, located at the surface of the nanobubble, exhibit fewer hydrogen bonds per molecule than Class IV molecules, which are closest to the bulk in terms of hydrogen bonding. Interestingly, we also observe that the $O \cdots H$ distance in hydrogen bonds is shorter for Class III molecules, showing that the hydrogen bonds are the strongest at the surface of the nanobubble. This is consistent with the experimental observations of Ohgaki {\it et al.}~\cite{Ohgaki}, who carried out attenuated total reflectance infrared spectroscopy to show that strong hydrogen bonds formed on the surface of nanobubbles. The simulation results presented here therefore shed light on two molecular processes that assist the formation and stability of nanobubbles, through the adsorption of vapor-like molecules close to the hydrophobic surface of the CNT and through the existence of strong hydrogen bonds between water molecules located at the vapor-liquid interface defining the nanobubble.

How does the extent of the nanoscopic confinement impact the nanobubble nucleation process? To address this question, we compare the free energy profiles obtained for different CNTs under the same conditions of chemical potential and temperature. The snapshots shown in Fig.~\ref{Fig4} provide a direct comparison between the nanobubbles obtained at the top of the free energy barrier for CNT $(20,20)$ (Fig.~\ref{Fig4}(a)) and CNT $(26,26)$ (Fig.~\ref{Fig4}(b)). The free energy profiles reveal that the free energy barrier of nucleation are of the same order, i.e. of $10\pm1$~$k_BT$ for CNT $(20,20)$ and of $12\pm1$~$k_BT$ for CNT $(26,26)$, with a similar mechanistic pathway followed for the formation of the nanobubbles. Both nanobubbles spread across the entirety of the nanotube, with a critical volume that is shown to exhibit more than a two-fold increase as the CNT diameter increases from about $27$~\AA~ for CNT $(20,20)$ to $35$~\AA~ for CNT $(26,26)$. The nanobubbles share the same qualitative features, i.e. are located on a flat part of the free energy profile, are associated with a positive value for $P_{//}$, confirming the crucial role played by these two favorable factors and are stabilized by the presence of vapor-like water molecules close to the hydrophobic surface and by the existence of strong hydrogen bonds at the vapor-liquid interface of the nanobubble. Future work will include the development of a theoretical model, linking the height of the free energy barrier to the contributions arising from the vapor-liquid surface tension, line tension and Tolman corrections~\cite{widom1995line,sharma2012evaporation,Duff}.

\section{Conclusions}

To shed light on the unexpected and mysterious stability of nanobubbles, we unravel in this work the nucleation pathway corresponding to the formation of nanobubbles in water confined in a carbon nanotube. We achieve this by using a molecular simulation method, function of an entropic order parameter, and by analyzing the ordering processes that takes place within the nanoconfined liquid and results in the formation of a nanobubble of a critical size. Several key factors accounting for the stabilization of such nanobubbles are identified. First, from a thermodynamic standpoint, configurations containing a nanobubble of a critical size are located on a flat part of the free energy profile. This leads to a prolonged stabilization of the nanobubble, since the absence of a strong free energy gradient will result in a very slow dissolution of the nanobubble. Second, the formation of this nanobubble occurs with a sign change in fluid pressure, which becomes positive again, leading to a mechanical stability for these nanobubbles. Third, from a structural standpoint, the stabilization of the nanobubble is assisted by the adsorption of vapor-like molecules close to the hydrophobic surface and by the onset of strong hydrogen bonds at the vapor-liquid interface, thereby confirming recent experimental and theoretical findings~\cite{Ohgaki,Zhang,Brenner,Weijs}. Our results also suggest how the stabilization of nanobubbles can be controlled through the choice of thermodynamic conditions in the absence of any additives, {\it e.g.}, ions to stabilize the vapor-liquid interface. This is key for many applications of nanobubbles, for energy production applications since nanobubbles can serve as high-pressure nanoreactors for fuell cell systems, and for biological systems, as nanobubbles provide a means for gas transport to membranes and cells.\\

{\bf Acknowledgements}
Partial funding for this research was provided by the National Science Foundation (NSF) through CAREER award DMR-1052808.\\

\bibliography{bubbles}

\end{document}